\documentclass[onecolumn,oneside]{jaa}

\usepackage{graphicx}

\def\ms{\ifmmode {\rm M_{\odot}} \else ${\rm M_{\odot}}$\fi}
\def\be{\begin{equation}}
\def\ee{\end{equation}}

\begin{document}\sloppy

\title{Probability Distribution of Magnetic Field Strengths through the Cyclotron Lines in High-Mass X-ray Binaries}


\author{Ali, Taani\textsuperscript{1,*}, Awni Khasawneh\textsuperscript{2}, Mohammad K.\ Mardini\textsuperscript{3,4}, Ahmad Abushattal\textsuperscript{5}, and Mashhoor Al-Wardat\textsuperscript{6,7,8}}
\affilOne{\textsuperscript{1}Physics Department, Faculty of Science, Al-Balqa Applied University,  Al-Salt 19117, Jordan.\\}
\affilTwo{\textsuperscript{2}Regional Centre for Space Science and Technology Education for Western Asia-UN, Amman, Jordan.\\}
\affilThree{\textsuperscript{3}Key Lab of Optical Astronomy, National Astronomical Observatories, Chinese Academy of Sciences, Beijing 100102, China.\\}
\affilFour{\textsuperscript{4}School of Astronomy and Space Science, University of Chinese Academy of Sciences, Beijing, 100049, China.\\}
\affilFive{\textsuperscript{5}Department of Physics, Al-Hussein Bin Talal University, 71111, Ma'an, Jordan.\\}
\affilSix{\textsuperscript{6}Department of Applied Physics and Astronomy, University of Sharjah, P. O. Box 27272 Sharjah, UAE.\\}
\affilSeven{\textsuperscript{7}Sharjah Academy for Astronomy, Space Science and Technology Sharjah, Sharjah,  UAE.\\}
\affilEight{\textsuperscript{8}Department of  Physics, Al al-Bayt University, Mafraq, 25113 Jordan.\\}


\twocolumn[{

\maketitle

\corres{ali.taani@bau.edu.jo}


\begin{abstract}

The study of variation of measured cyclotron lines is of fundamental importance to understand the physics of the accretion process in magnetized neutron star systems. We investigate the magnetic field formation, evolution and distribution for several High- Mass X-ray Binaries (HMXBs). We focus our attention on the cyclotron lines that have been detected in HMXB classes in their X-ray spectra. As has been correctly pointed out, several sources show variation in cyclotron lines, this can result due to the effect of accretion dynamics, and hence that would reflect the magnetic field characteristics. Besides, the difference in time scales of variation of accretion rate and different type of companion can be used to distinguish between magnetized neutron stars.

\end{abstract}


}]


\setcounter{page}{1}

\section{Introduction}

The cyclotron lines in accreting neutron stars (NSs) are detected as absorption features in their X-ray spectra. This detection is due to resonant scattering processes with electrons which are perpendicular to the B-field (Truemper et al. 1978; Voges et al. 1982; Wilson et al. 2008; Taani et al. 2019a,b). The significant detection of such behavior helps us to study the formation and evolution of X-ray binaries through the direct calculation of the magnetic field strength. However, most observed cyclotron lines have been detected above 10 keV and are interpreted as electron features, with inferred magnetic fields B~ 10$^{12}$ G (Heindl et al. 2001). The amount of mass loss and the effect of the mass-loss via stellar winds may influence the stellar binary evolution. This will also allow distinguishing the phenomenology of the X-ray sources and their optical counterpart in a natural way (Taani et al. 2012, 2019a,b; Shakura et al. 2012; Taani 2016; Mardini et al 2019a,b,c; Karino et al. 2019). Despite its importance, many questions remain unanswered in terms of the accretion geometry and flow with significant uncertainties due to the small number of detected cyclotron lines in some sources (Coburn et al. 2002; Kreykenbohm et al. 2005; Taani et al. 2013; Taani $\&$ Khasawneh 2017). In addition, a cutoff below 2-3 keV would remain undetected in the available spectra of some sources. Therefore, the physical properties of the accretion column, as well as the line profiles of cyclotron lines must be studied in detailed. Nishimura (2005) calculated cyclotron lines assuming a strong variation on field strength with a distance from an emission region. However, no model generating such high flux and high temperature at a layer deeper than absorbing heavy atoms has been proposed. According to recent studies, several pulsars show the luminosity dependence changes in the cyclotron resonance energy. The main aim of this paper is to investigate the probability distribution of magnetic field strengths among the HMXBs, using a more robust values of the B-field obtained from cyclotron lines in their X-ray spectra (see Table 1). One of the interesting properties of this class is the observed correlation between the orbital period and the spin period of the NS (Corbet 1986). This would reveal the clues about the evolution of HMXBs, which can be understood in terms of the conservative evolution of normal massive binary systems.

\section{STATISTICAL TESTS}

We make use of two statistical tests to try to quantitatively evaluate our sample of magnetic field strength (see Table 1): the Kologorov-Smirnov (KS) test and the Anderson- Darling (AD) test (e.g., Press et al. 2007; Cai et al. 2012). Because the AD test uses the cumulative distribution, we use it as our primary comparison and use the KS test results as a consistency check. The first test we perform is for normality, checking whether the distribution is consistent with a single Gaussian which has the mean and standard deviation of the observed populations. Neither the mean nor the variance is known beforehand for the distributions, and we test at the 5$\%$ level of significance. Furthermore, we use a Monte-Carlo (MC) to test the confidence that the given distribution is either normally or log-normally distributed. Our results are tabulated in Table 2. The results of the KS test agree in every case with those of the AD test and have therefore been omitted from the table. From the table, it is clear that every distribution, with the exception of the magnetic field for the transient sources, fails the AD test for normality. However, every distribution passes the test for lognormality. The confidence from the MC tests is all below 50$\%$ for the persistent sources and slightly above 50$\%$ for the spin and orbit distributions of the transient sources. While these statistical tests and fits do not, by themselves, constitute a proof of two populations, coupled with the other pieces of evidence (i.e., such as mass transfer due to Roche lobe overflow or stellar wind that can also be a supplement to the mass transfer rate in HMXB systems, tidal interaction, gravitational wave radiation, magnetic braking or X-ray irradiated wind outflow), they do lend support for the hypothesis of two populations. We show in Fig 1, the cumulative distribution function of the energy of cyclotron absorption line (left) and magnetic field strength (right) of the observed sample HMXBs. We note that this function is not smooth, implying that the energy of the lines are not constant but change linearly with the luminosity of the sources (Tsygankov et al. 2012). We excluded the data after 65 KeV, because the high field cut-off appears to be real. On the observational level, this variability in transient sources, for example, is most likely due to the irregular optical and IR outbursts generally observed in Be stars, and it is attributed to changes in the presence structure of the circumstellar disk. These effects were investigated by Reig et al. (2001).

\begin{figure*}[h]
\centering
\includegraphics[width=\textwidth]{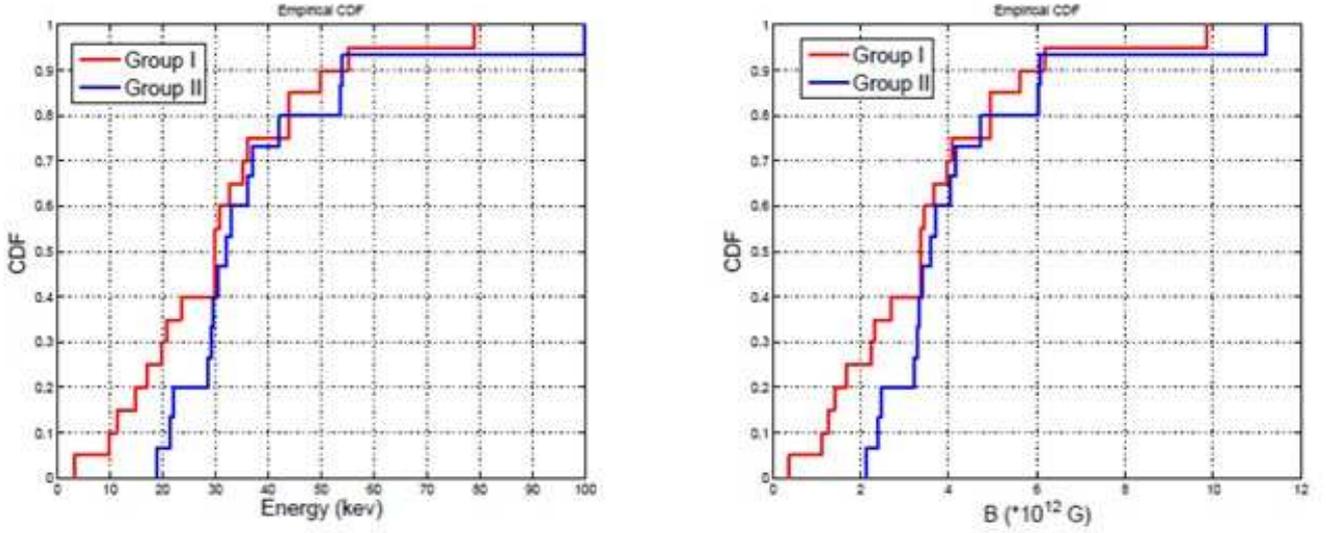}
\caption{The cumulative distribution function of cyclotron energy and magnetic field strength of the observed sample HMXBs, dividing them into two groups, transient sources (Group I) and persistent sources (Group II).}
\label{Fig2}
\end{figure*}

\section{\textbf{P}$_{SPIN}$ Vs. \textbf{P}$_{ORB}$}

The \textbf{P}$_{SPIN}$ versus \textbf{P}$_{ORB}$ diagram (also known as the Corbet diagram, see Fig. 2) (Corbet 1986) is a valuable tool to study the interaction and feedback between the NS and accreted matter, and the influence of the local absorbing matter, the location of the different systems being determined by the equilibrium period reached by the rotation of the NS accreting matter on its surface. The main parameters of the HMXB sample are summarized in Table 1. However, the orbital period of X-ray binaries is expected to change due to redistribution of the angular momentum due to the interaction between the components of the binary system. As such, measurement of the rate of change of the orbital period (i.e., orbital period derivative) of the binary system is, therefore, necessary in order to understand the evolution of compact binary systems (Rivers et al. 2010; Wei et al. 2010). The spin and orbital periods of all HMXBs for which values are known are plotted in Fig. 2, which represents a panoramic perspective for binaries and occupy separate parts of the plot. This is not only because the NSs in HMXBs have a different type of companion, but also because the accretion process itself seems to be universal (Taani et al. 2019a,b), with the NS spin in or near an equilibrium state in which the magnetospheric radius of the NS equals the Keplerian co-rotation radius. They are a group of supergiant (SG) sources having peculiar properties of orbital parameters and could be a good test-candidate for those in Be sources. Future observations can identify these candidates. This group is categorized according to their observed spin period as in table 1. It is noteworthy to mention here that, the effect of mass exchange process on the orbital evolution has two scenarios 1) conservative case: where the total mass and the total angular momentum of the system do not change (see i.,e. Dosopoulou $\&$ Kalogera 2016; Bobrick et al. 2017). Hence, the size and the orbital period of the system must be decreased if the mass is transferred to the less massive component (NS) and vice-versa. 2) non conservative case (more complex): this will depend on how and how much angular momentum is lost from the system (see i.,e. by Demircan et al. 2006; Eker et al. 2006 for full details and references).

It may be interesting to investigate the relation between the magnetic field strength and spin period, also known as the spin-up line, (Bhattacharya $\&$ van den Heuvel 1991) to discuss the formation and evolution of these systems through various evolutionary stages. In Fig. 3, it represents a plot of the magnetic field strength as a function of the spin period in logarithmic scale and cyclotron energy colored. It can be seen clearly that there is a possible correlation between the magnetic field strength and the spin period. Moreover, the cyclotron energy independence of the spin period is very clear. This implies that the NSs in these systems generally have ages $\sim 10^{6}$ years. However, the presence of strong magnetic fields (B $>$ $10^{12}$ G) in the HMXBs as evidenced by their regular x-ray pulsations, while the absence of such regular pulsations and thus strong magnetic fields in the Low Mass X-ray Binaries (LMXBs) (van den Heuvel 2004). However, the known magnetic field strengths of the X-ray pulsars are all lying in a very narrow band (due to the observational selection effect) and can be used by the equation

\be
\label{E1}
E_{cyc} = 11.6 \times B_{12}\times(1+Z)^{-1}\;,
\ee

where z is the gravitational redshift at the NS surface is approximately (z = 0.3) in the line forming region (Wasserman $\&$ Shapiro 1983). Here B$^{12}$ is the magnetic field strength in units of 10$^{12}$ G, and the higher harmonics have an cyclotron energy n times the fundamental energy E$_{cyc}$. Using equation (1), the surface magnetic field strengths of HMXBs sample have been calculated and presented in table 1. We found that these values are clustered in a relatively narrow range of (1-13.3) $\times$ 10$^{12}$ G.

\begin{figure}[h]
\centering
\includegraphics[width=0.47\textwidth]{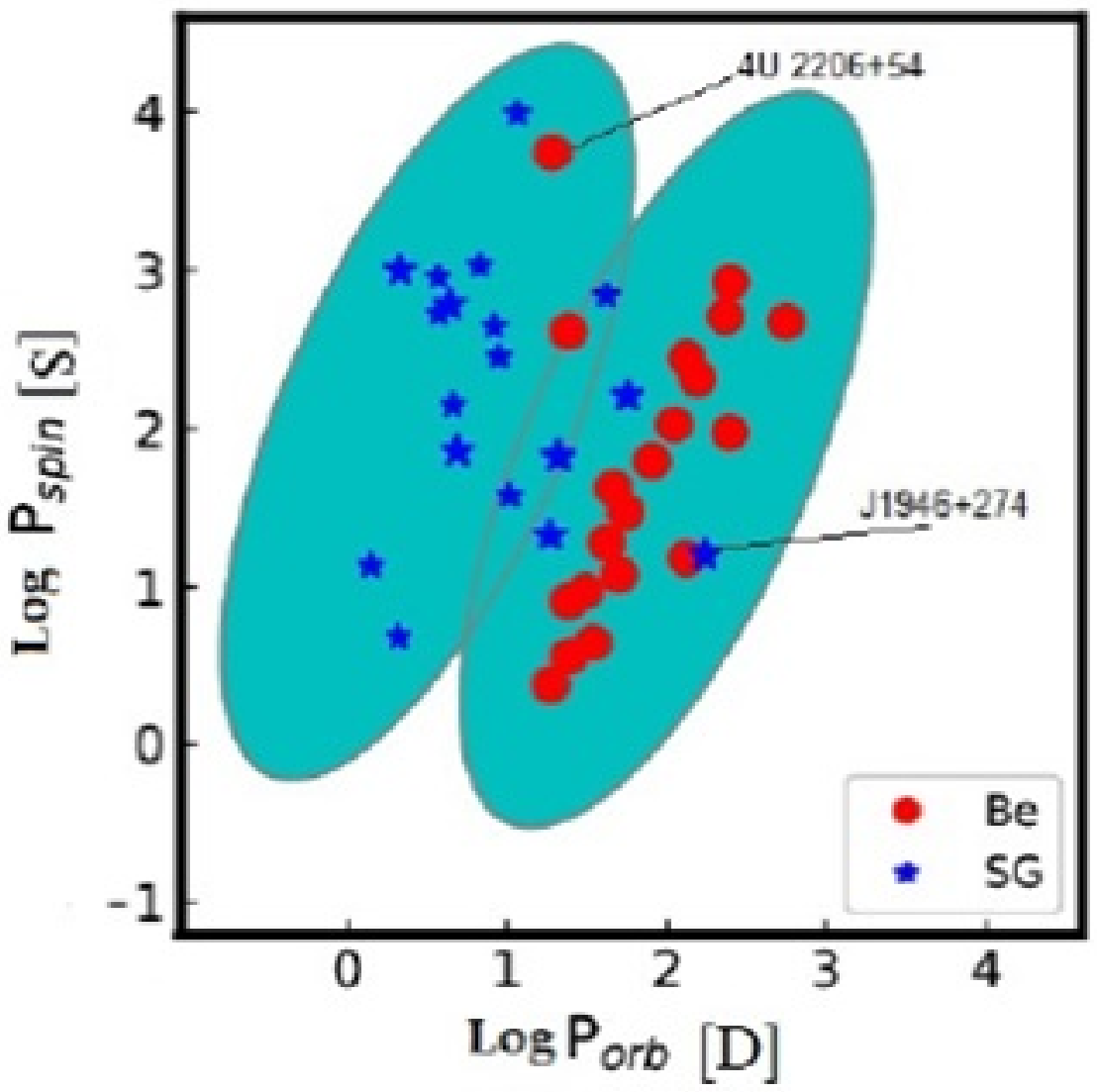}
\caption{Corbet Diagram for all the HMXBs that have presented of a cyclotron line. A group of SG and Be clusters that act together and occupy the same regions of the parameter space; akin to the separated regions on this figure. Note that the peculiar properties of J1946+274 and 4U 2206+54 are let them be observed in Be and SG regions respectively. Future observation and the modeling including disk/wind fed with orbital motion would figure out the true nature and any further evolution for such binaries. The shaded regions represent the potential observations of the orbital parameters those groups.
}
\label{Fig2}
\end{figure}

\newcounter{rit} 
\newcommand{\rit}{\refstepcounter{rit}\therit}

\begin{table*}
\caption{List of observational parameters of all known of HMXBs with
cyclotron resonant scattering features}
\label{O-I}
\setlength{\tabcolsep}{0.001pt}
\renewcommand{\arraystretch}{0.001}
\begin{tabular}{lcccccccl}
\hline \hline \noalign{\smallskip}
 \multicolumn{1}{c}{Object} &
\multicolumn{1}{c}{$P_{\rm spin}$} & \multicolumn{1}{c}{$P_{\rm
orbit}$} & \multicolumn{1}{c}{$E_{\rm cyc}$} &
\multicolumn{1}{c}{$B$} &
\multicolumn{1}{c}{Type} &
\multicolumn{1}{c}{Ref.} \\
\multicolumn{1}{c}{}& \multicolumn{1}{c}{(s)} & \multicolumn{1}{c}{(d)} &
\multicolumn{1}{c}{(keV)} &
\multicolumn{1}{c}{(10$^{12}$G)}   &
\multicolumn{1}{c}{}\\
\hline \noalign{\smallskip}
\\ 4U 0115$+$63 & 3.6 & 24.3 & 15$\pm$0.15
                      & 1.7 &  Transient/ Be &
                      \ref{1979Wheaton}, \ref{1991Nagase}, 
                      \ref{2008Wilson}, \ref{2011Ferrigno}\\
4U 1907$+$09 & 439 & 8.37 & 18.8$\pm$0.4  &2.1&Persistent/
SG &              \ref{1998Cusumano}, 
              \ref{2002Coburn}, \ref{2010Rivers}\\
4U 1538$-$52 & 529 & 3.73
             & 21.4$^{+0.9}_{-2.4}$  &2.4& Persistent/SG &
             \ref{2002Coburn}, \ref{1990Clark}, 
              \ref{2001Robba},  \ref{2009Rodes-Roca} \\
Vela X-1 & 283 & 8.96 & 54$^{+0.5}_{-1.1}$ &  6&Persistent/ SG  &
              \ref{1996Kretschmar}, \ref{1999Makishima}, \ref{2002Kreykenbohm}, \ref{2007Schanne}\\
Cen X-3 & 4.8 & 2.09 & 30.4$^{+0.3}_{-0.4}$ & 3.4 &
Persistent/ SG &         \ref{2002Coburn}, \ref{1999Makishima}, \ref{1998Santangelo}  \\
V0332+53 & 4.37 & 34.25
                     & 30$^{+0.2}_{-0.2}$  &3.4 &Transient/ Be &
                      \ref{1990Makishima}, 
                      \ref{2005Kreykenbohm}, \ref{2005Pottschmidt}, \ref{2010Nakajima}\\
Cep X-4 & 66.3 & 20.85 & 30.7$^{+1.8}_{-1.9}$ &3.4&Transient/ SG &        \ref{1999Makishima}, \ref{1991Mihara},  \ref{2007McBride} \\
A 0535$+$26 & 105 & 111 & 50$\pm$0.7 &  5.6& Transient/
Be&             
             \ref{2008Wilson}, \ref{1996Kretschmar}, 
             \ref{2005Wilson}, \ref{2006Terada}  \\
GX 301$-$2 & 690 & 41.5 &42.4$^{+3.8}_{-2.5}$ & 4.7& Persistent/ Be& \ref{2002Coburn}, \ref{1999Makishima} \\
LMC X-4 & 13.5 & 1.4 & 100$\pm$2.1 & 11.2 & Persistent/
SG&
         \ref{1999Makishima}, \ref{2001Barbera} \\
4U 0352$+$309 & 837 & 250 & 28.6$^{+1.5}_{-1.7}$  & 3.2&Persistent/ Be&
              \ref{2002Coburn}, \ref{2001Coburn} \\

 OAO1657-415 & 37.7& 10.4& 36&  4& Persistent/
SG&
 \ref{1999Orlandini}, \ref{2010Denis}, \ref{2011Pottschmidt}\\
J1946+274 & 15.83 & 169.2 & 36.2$^{+0.5}_{-0.7}$  &4&Transient/SG&             \ref{2002Coburn}, \ref{2001Heindl}, \ref{2003Wilson} \\
MXB 0656$-$072 & 160.4 & 56.2 & 32.8$^{+0.5}_{-0.4}$  &3.7& Transient/SG &                \ref{2003Heindl}, \ref{2006McBride} \\
GX 304$-$1 & 275.46 & 132.5 & 53.7$^{+0.7}_{-0.6}$  &6&Transient/Be &  \ref{2010Sakamoto},
            \ref{2011Yamamoto} \\
J16493-4348$^{\dag}$ & 1069 & 6.78 & 33$\pm$4  & 3.7 &Persistent/SG & \ref{2010Nespoli}, \ref{2011D'Ai}  \\
GS 1843+00 & 29.5  &  55 & 20$\pm$0.45 &  2.2& Transient/Be& \ref{2000Piraino}, \ref{2011Pottschmidt}  \\
1A1118-61 & 408 & 580 & 55.1$^{+1.6}_{-1.5}$ & 6&Transient/Be& \ref{2010Doroshenko}, \ref{2011Devasia}, \ref{2011Nespoli} \\
J1008-57 &  93.5 &   247.8 & 79 & 10&Transient/Be& \ref{2006Wilms}, \ref{2011Naik}, \ref{2014Sguera}\\
EXO 2030+375 &41.7 & 46 & 11.44$\pm$0.02&  1.3& Transient/Be& \ref{2008Wilson}, \ref{2010Sasaki}\\
J1626.6-5156 & 15& 132& 10 &  1.1 & Transient/Be&  \ref{2010Baykal}, \ref{2009deCesar} \\
4U 1700-377 & --&3.4& 37&  4.1& Persistent/SG& \ref{2011Pottschmidt}, \ref{1999Reynolds}\\
J01583+6713 &469&561&$35.3\pm1.6$& 4& Transient/Be & \ref{2010Wang}\\
4U 2206+54 & 5500& 19.11& $29.6\pm2.8$&  3.3& Persistent/Be&  \ref{2009Wang}, \ref{2009Tomsick}\\
2S 0114+65& 9700& 11.6& 22&  2.5& Persistent/SG&
\ref{2005Bonning}, \ref{2006denHartog}\\
J1739-302$^{\dag}$ &--&51.5& 30&  3.4&   Transient/SG& \ref{2008Blay}, \ref{2010Drave}\\
J18483-0311$^{\dag}$& 21&18.6&3.3&0.4&Transient/SG& \ref{2003Chernyakova}, \ref{2010Sguera} \\
J0440.9+4431& 205& 155& 32&3.6& Persistent/Be&\ref{2012Tsygankov}\\
J1409 - 619 &506&233& 44$\pm$3&4.9&Transient/Be&\ref{2012Orlandini}\\
J18462-0223&997 &--&30$\pm$7&3.4&Transient/SG&\ref{2010Grebenev}\\
J18179-1621&11.82&20-50&20.8$^{+1.4}_{-1.8}$&2.3&Transient/Be&\ref{2012Tuerler}\\
J17544-261 &71.5 & 4.9 & 17 &  1.45 &  Transient/SG& \ref{2015Bhalerao}\\
2S 1553-542&9.27&30.6&23.5$\pm$0.4&2.7&Transient/Be&\ref{2015Tsygankov}\\
4U 1909+07&604&4.4&44&4.9&Transient/SG& \ref{2013Jaisawal}\\
J16393-4643&904&4.2&29.3$^{+1.1}_{-1.3}$&3.3&Persistent/SG& \ref{2016Bodaghee}\\
J054134.7-68&61.6&80&10&1.2&Persistent/Be& \ref{2009Manousakis}\\
KS 1947+300&1808&41.5&12.5&1.4&Transient/Be& \ref{2014Furst}\\
IGR J18027-201&140&4.6&23&2.6&Persistent/SG& \ref{2017Lutovinov}\\
SMC X-2&2.4&18.6&27&3.1&Transient/Be& \ref{2016Jaisawal}\\
J0520.5-69&8&24&31.5&3.6&Transient/Be& \ref{2014Tendulkar}\\
\hline \noalign{\smallskip}
\end{tabular}

  $^{\dag}$: Candidate sources for cyclotron features, References.-- These references are to period measurements in the literature.
  Some have errors originating from applied analysis, designated
with a dagger, or from the supplied data, designated with an
asterisk. (\rit\label{1979Wheaton}) Wheaton et al. 1979;
(\rit\label{1991Nagase}) Nagase et al. 1991;
(\rit\label{2008Wilson}) {Wilson} et al. 2008
(\rit\label{2011Ferrigno}) {Ferrigno} et al. 2011
 (\rit\label{1998Cusumano}) Cusumano et al. 1998;
 (\rit\label{2002Coburn}) Coburn et al. 2002;
 (\rit\label{2010Rivers}) Rivers et al. 2010;
 (\rit\label{1990Clark}) Clark et al. 1990;
 (\rit\label{2009Rodes-Roca}) Rodes-Roca et al 2009;
 (\rit\label{2001Robba}) Robba et al. 2001;
 (\rit\label{1996Kretschmar}) Kretschmar et al. 2005;
  (\rit\label{1999Makishima}) Makishima et al. 1999;
  (\rit\label{2002Kreykenbohm}) Kreykenbohm et al. 2002
  (\rit\label{2007Schanne}) Schanne  et al. 2007
  (\rit\label{1998Santangelo}) Santangelo et al. 1998;
  (\rit\label{1990Makishima}) Makishima et al. 1990;
  (\rit\label{2005Kreykenbohm}) Kreykenbohm et al. 2005;
  (\rit\label{2005Pottschmidt}) Pottschmidt et al. 2005;
  (\rit\label{2010Nakajima}) Nakajima et al. 2010;
  (\rit\label{1991Mihara}) Mihara et al. 1991;
  (\rit\label{2007McBride}) McBride et al. 2007;
  (\rit\label{1978Trumper}) Tr\"{u}mper et al. 1978;
  (\rit\label{1982Voges}) Voges et al. 1982;
(\rit\label{2008Enoto}) Enoto et al. 2008;
 (\rit\label{2008Klochkov}) Klochkov et al. 2008;
(\rit\label{2005Wilson}) Wilson \& Finger 2005;
(\rit\label{2006Terada}) Terada et al. 2006;
(\rit\label{2001Barbera}) Barbera et al. 2001;
(\rit\label{2001Coburn}) Coburn et al. 2001;
(\rit\label{1999Orlandini}) Orlandini et al. 1999;
(\rit\label{2010Denis}) Denis et al. 2010;
(\rit\label{2011Pottschmidt}) {Pottschmidt}, S et al. 2011
  (\rit\label{2001Heindl}) Heindl et al. 2001;
  (\rit\label{2003Wilson}) {Wilson} et al. 2003
(\rit\label{2003Heindl}) Heindl 2003;
  (\rit\label{2006McBride}) McBride et al 2006;
  (\rit\label{2010Sakamoto}) Sakamoto et al. 2010;
  (\rit\label{2011Yamamoto}) Yamamoto et al. 2011;
  (\rit\label{2010Nespoli}) {Nespoli} et al. 2010
  (\rit\label{2011D'Ai}) {D'Ai} et al. 2011
  (\rit\label{2000Piraino}) {Piraino} et al. 2000
  (\rit\label{2010Doroshenko}) {Doroshenko} et al. 2010
  (\rit\label{2011Devasia}) {Devasia} et al. 2011
(\rit\label{2011Nespoli}) {Nespoli} et al. 2011
    (\rit\label{2006Wilms}) {Wilms}  2006
  (\rit\label{2011Naik}) {Naik} et al. 2011
  (\rit\label{2014Sguera}) {Sguera} et al. 2014
(\rit\label{2010Sasaki}) {Sasaki}  2010
(\rit\label{2010Baykal}) {Baykal} et al. 2010
(\rit\label{2009deCesar}) {deCesar} et al. 2009
(\rit\label{1999Reynolds}) {Reynolds} et al. 1999
(\rit\label{2010Wang}) {Wang} et al. 2010
(\rit\label{2009Wang}){Wang} et al. 2009
(\rit\label{2009Tomsick}){Tomsick} et al. 2009
(\rit\label{2005Bonning}) {Bonning} et al. 2005
(\rit\label{2006denHartog}) {denHartog} et al. 2006
(\rit\label{2008Blay}) {Blay} et al. 2008
\rit\label{2010Drave}){Drave} et al. 2010
\rit\label{2003Chernyakova}) {Chernyakova} et al. 2003
\rit\label{2010Sguera}) {Sguera} et al. 2010
(\rit\label{2012Tsygankov}) {Tsygankov} et al. 2012.~
(\rit\label{2012Orlandini}) Orlandini et al. 2012;
(\rit\label{2010Grebenev}) Grebenev\&Grebenev 2010;
(\rit\label{2012Tuerler}) Tuerler et al. 2012;
(\rit\label{2015Bhalerao}) Bhalerao et al. 2015;
(\rit\label{2015Tsygankov}) {Tsygankov} et al. 2015.~
(\rit\label{2013Jaisawal}) {Jaisawal} et al. 2013.~
(\rit\label{2016Bodaghee}) {Bodaghee} et al. 2016.~
(\rit\label{2009Manousakis}) {Manousakis} et al. 2009.~
(\rit\label{2014Furst}) {Furst} et al. 2014.~
(\rit\label{2017Lutovinov}) {Lutovinov} et al. 2017.~
(\rit\label{2016Jaisawal}) {Jaisawal} et al. 2016.~
(\rit\label{2014Tendulkar}) {Tendulkar} et al. 2014.~
\end{table*}

\begin{figure}[h]
\centering
\includegraphics[width=0.47\textwidth]{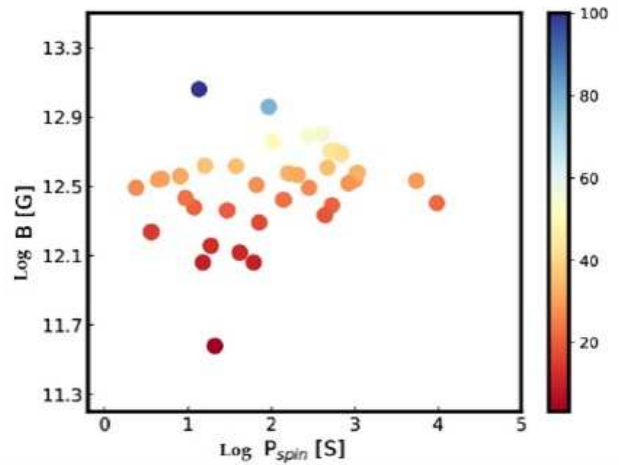}
\caption{The diagram of the magnetic field strengths versus spin periods for our data sample of HMXBs (SG and Be sources) in Table 1. Color-coded are based on the associated cyclotron energy. Sources with the same energy (same color) show zero slope against the spin period is appeared.
}
\label{Fig2}
\end{figure}

\begin{figure}[h]
\centering
\includegraphics[width=0.47\textwidth]{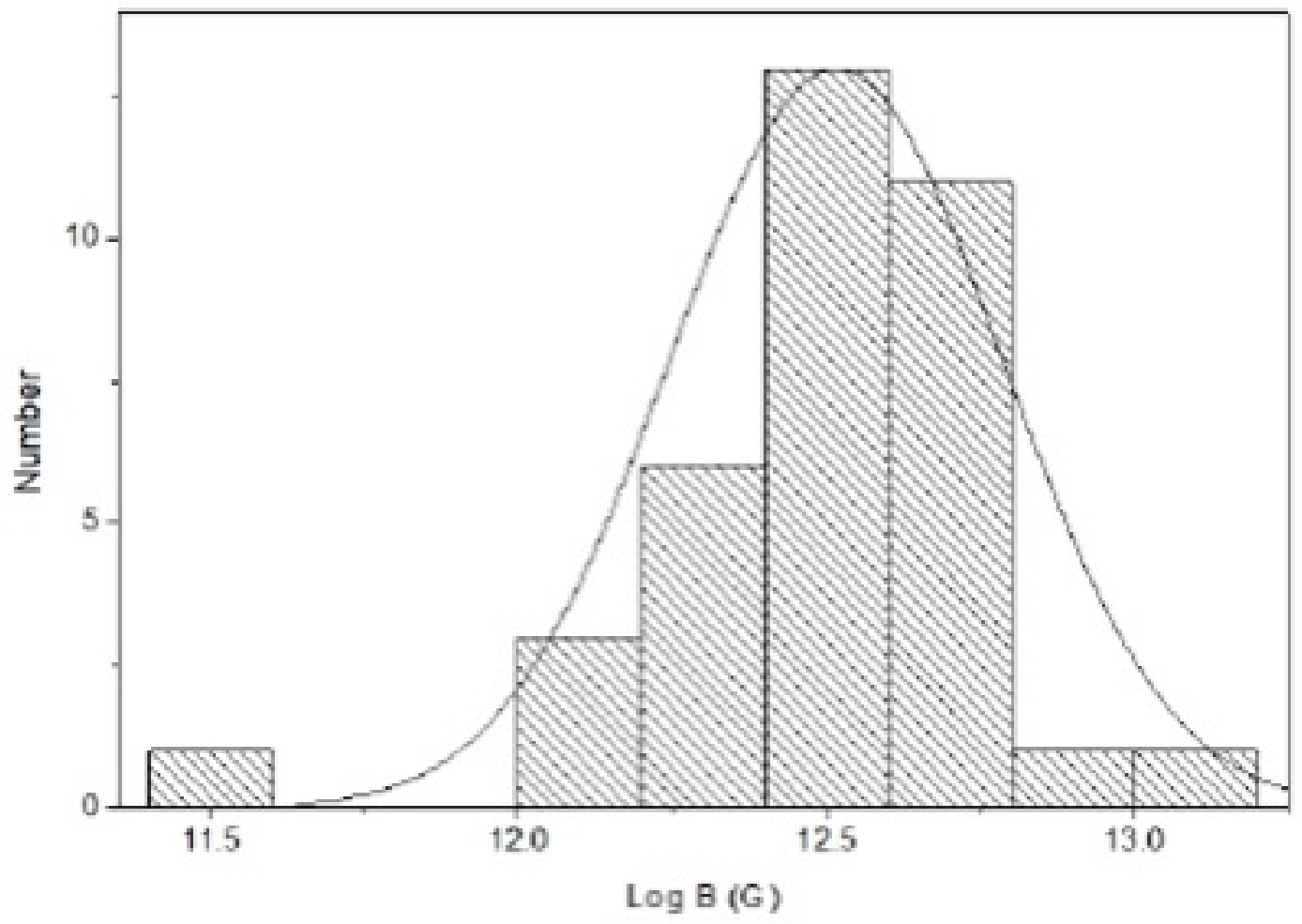}
\caption{Magnetic field strength distribution of the observed sample. The solid line is the curve fitted using a Gaussian function.}
\label{Fig2}
\end{figure}

To illustrate all HMXB magnetic field strength distributions, a histogram of HMXB magnetic field strengths is plotted in Fig. 4, and the smooth curve represents the normal fit to the observed data. We assume that the distribution follows a Gaussian distribution, centered at 12.5 $\times$ 10$^{12}$ G. In these systems, the data suffer from selection effect and may be due to a sensitivity of the current X-ray observatories. About 89$\%$ of all HMXBs, considered in this study, has high magnetic field B = 12.5 $\times$ 10$^{12}$ G. The maximum and minimum values of HMXB magnetic fields are, respectively, 11.2 $\times$ 10$^{12}$ G (J0520.5-69) and 1.1 $\times$ 10$^{12}$ G (J1626$-$5156). We fit the Gauss function to the magnetic distributions. The Gauss function we choose reads,

\be
\label{E1}
y = y_{0}+ \frac{A}{w\sqrt{\frac{\pi}{2}}} exp~(-2 \times (\frac{x-x_{c}}{w})^{2})\;,
\ee
here, the parameters y$_{0}$, x$_{c}$, w and A are offset of vertical axis, center of horizontal axis, width and area the curve describing, respectively. The fitting results are listed in Table 3.

\section{SUMMARY AND CONCLUSIONS}

The magnetic field strength of several HMXB systems has been investigated. As such, the magnetic field is responsible for the formation of an accretion column since it forces the particles to hit the NS surface at its magnetic poles. Therefore, the cyclotron lines are of fundamental importance to understand the properties of other physical parameters of the magnetized X-ray system. In the picture presented here, we find that most NSs in HMXBs show an intermediate level of field strength, $\sim$ 10$^{12}$ G, however the different characteristics of their X-ray behaviors and properties. We analyze and discuss the distribution of the characteristics of the HMXB sample, through the benefit from the Corbet diagram. In particular one can see that all transient sources have an orbital period more extended than 11 d, thus newly detected sources for which P > P$_{11d}$ have a high probability of being transients. An attempt is made to quantify the possible distribution of the magnetic field strength through the cumulative probability distribution. The extent to which the sample is complete is well demonstrated by a log-normal law and shows somewhat similar behavior for both groups. Such parameters greatly affect the model of wind-fed binary systems and can be constrained by stellar properties during the binary evolution. However, the quantity of these exciting objects has greatly increased in recent years, mainly due to successful surveys. More details and results on the multiplicity of the cyclotron line features from some sources are required, to explain the behavior of shape variation during the rotation phase as well as the change in accretion rate and characteristics.

\begin{table*}
\caption[]{Fit parameters of the distribution of magnetic field strengths.}
\label{table:1}
\setlength{\tabcolsep}{2pt}
\begin{tabular}{lccccccr}
\hline
\hline
Quantity  &y$_{0}$  & X$_{c}$  & w   &A &$\sigma$ & $x^{2}/Dof$     &R$^{2}$ \\
\hline
Magnetic field     &12.6$\pm$0.43    &15.03$\pm$0.02     &2.38$\pm$0.02          &12.72$\pm$0.02         &0.22     &0.5    &$0.26$      \\
\end{tabular}
\end{table*}

\section*{Acknowledgements}

Ali Taani gratefully acknowledges the Institute of High Energy Physics, Chinese Academy of Sciences through the CAS-PIFI Fellowship 2018, where this work started. M. Mardini thanks the National Natural Science Foundation of China for the support under the grant numbers 11988101 and 11890694 and also the National Key R$\&$D Program of China No. 2019YFA0405500. A. Abushattal acknowledges the support from Al-Hussein Bin Talal University, Deanship of Scientific Research and Graduate Studies under grant No. (17/2018). The authors would like to thank the anonymous referees for helpful suggestions and comments on the manuscript.

\vspace{-1em}




\begin{theunbibliography}{}
\vspace{-1.5em}

\bibitem{latexcompanion}
Barbera A., Burderi L., Di Salvo T. et al. 2001, ApJ, 553, 375
\bibitem{latexcompanion}
Bhalerao V., Romano P., Tomsick T. $\&$ et al. 2015, MNRAS, 447, 2274
\bibitem{latexcompanion}
Bhattacharya D. $\&$ van den Heuvel E.J., 1991, Phys. Rep. 203, 1
\bibitem{latexcompanion}
Blay P., Martınez-Nunez S., Negueruela I. $\&$ et al. 2008, A$\&$A, 489,669
\bibitem{latexcompanion}
Bobrick A., Davies M. B., $\&$ Church R. P., 2017, MNRAS, 467, 3556
\bibitem{latexcompanion}
Bodaghee A, Tomsick J. A., Fornasini F. A. $\&$ et al. 2016, ApJ, 823, 146
\bibitem{latexcompanion}
Bonning E. W., $\&$ Falanga M., 2005, A$\&$A, 436, L31
\bibitem{latexcompanion}
Cai Y., Taani A., Zhao Y., C Zhang, 2012, Chinese Astronomy and Astrophysics, 36, 137
\bibitem{latexcompanion}
Chernyakova M. et al., 2003, Astron. Telegram, 157
\bibitem{latexcompanion}
Clark G. W., Woo J. W., Nagase F. $\&$ et al. 1990, ApJ, 353, 274
\bibitem{latexcompanion}
Coburn W., Heind W., Gruber D. et al. 2001, ApJ, 552, 738
\bibitem{latexcompanion}
Coburn W., Heind W., Rothschild R. et al. 2002, ApJ, 580, 394
\bibitem{latexcompanion}
Corbet R.H.D., 1986, MNRAS, 220, 1047
\bibitem{latexcompanion}
Cusumano G., de Salvo T., Burderi L. $\&$ et al. 1998, A$\&$A, 338, L79
\bibitem{latexcompanion}
D'Ai A., Cusumano G., La Parola V. et al. 2011, A$\&$A, 532, A73
\bibitem{latexcompanion}
DeCesar M. E., Pottschmidt K., $\&$ Wilms J., 2009, ATel, 2036
\bibitem{latexcompanion}
Demircan O., Eker Z., Karatas Y. et al. 2006, MNRAS, 366, 1511
\bibitem{latexcompanion}
Denis M., Bulik T. $\&$ Marcinkowski R., 2010, AcA, 60, 75D
\bibitem{latexcompanion}
den Hartog P. R., Hermsen W., Kuiper L. $\&$ et al. 2006, A$\&$A, 451, 587
\bibitem{latexcompanion}
Devasia J., James M., Paul B. et al. 2011, MNRAS, 414, 1023
\bibitem{latexcompanion}
Doroshenko V., Suchy S., Santangelo A. et al. 2010, A$\&$A, 515, L1
\bibitem{latexcompanion}
Dosopoulou F. $\&$ Kalogera V., 2016, ApJ, 825, 70
\bibitem{latexcompanion}
Drave S. P., Clark D. J., Bird A. J. $\&$ et al. 2010, MNRAS, 409, 1220
\bibitem{latexcompanion}
Eker Z., Demircan O., Billir S. et al., 2006, MNRAS, 373, 1483
\bibitem{latexcompanion}
Enoto T., Makishima K., Terada Y. et al. 2008, PASJ, 60, 57
\bibitem{latexcompanion}
Ferrigno C., Falanga M., Bozzo E. et al. 2011, A$\&$A 532, A76
\bibitem{latexcompanion}
Furst F., Pottschmidt K., Wilms J. et al., 2014, ApJ, 784, L40
\bibitem{latexcompanion}
Heindl W., Coburn W., Gruber D. et al. 2001, ApJ 563, L35
\bibitem{latexcompanion}
Heindl W., Coburn W., Kreykenbohm I. et al. 2003, ATel, 200, 1
\bibitem{latexcompanion}
Grebenev S.A. $\&$ Sunayev R.A., Astr. Lett, 2010, 36, 53
\bibitem{latexcompanion}
Jaisawal G. K., Naik S., Paul B. $\&$et al. 2013, ApJ, 779, 54
\bibitem{latexcompanion}
Jaisawal G. $\&$ Naik S. 2016, MNRAS, 461, L97
\bibitem{latexcompanion}
Karino S., Nakamura K $\&$ Taani A., 2019, Publications of the Astronomical Society of Japan, 71, 58
\bibitem{latexcompanion}
Klochkov D., Staubert R., Postnov K. $\&$ et al. 2008, A$\&$A, 482, 907
\bibitem{latexcompanion}
Kretschmar P., Kreykenbohm I., Pottschmidt K. $\&$ et al. 2005, ATel, 601
\bibitem{latexcompanion}
Kreykenbohm I., Coburn W., Wilms J. et al. 2002, A$\&$A, 395, 129
\bibitem{latexcompanion}
Kreykenbohm I., Mowlavi N., Produit N. et al. 2005, A$\&$A, 433, 45
\bibitem{latexcompanion}
Lutovinov, A., Tsygankov, S. 2017, MNRAS, 466
\bibitem{latexcompanion}
Makishima K., Mihara T., Ishida M., 1990, ApJ, 365, 59
\bibitem{latexcompanion}
Makishima K., Mihara T., Nagase F. et al. 1999, ApJ, 525, L97
\bibitem{latexcompanion}
Manousakis A., Walter R., Audard M. et al. 2009, A$\&$A, 498, 217
\bibitem{latexcompanion}
Mardini, M., Ershiadat N., Al-Wardat M. 2019, Journal of Physics: Conference Series, 1258, 12024
\bibitem{latexcompanion}
Mardini, M. K., Placco, V. M., Taani, A., Li, H., $\&$ Zhao, G. 2019b, ApJ, 882, 27
\bibitem{latexcompanion}
Mardini, M. K., Li, H., Placco, V. M., et al. 2019, ApJ, 875, 89
\bibitem{latexcompanion}
McBride V., Wilms J., Coe M. J. et al. 2006, A$\&$A, 451, 267
\bibitem{latexcompanion}
McBride V., Wilms J., Kreykenbohm I. et al. 2007, A$\&$A, 4470, 1065M
\bibitem{latexcompanion}
Mihara T., Makishima K., Ohashi T. et al. 1990, Nature, 346, 250
\bibitem{latexcompanion}
Naik S., Paul B., Kachhara C. et al. 2011, MNRAS, 413, 241
\bibitem{latexcompanion}
Nagase F., Dotani T., Tanaka Y. et al. 1991, ApJ, 375, 49
\bibitem{latexcompanion}
Nakajima M., Mihara T., Makishima K., 2010, ApJ, 710, 1755
\bibitem{latexcompanion}
Nishimura O., 2005, PASJ, 57, 769
\bibitem{latexcompanion}
Nespoli E., Fabregat J., $\&$ Mennickent R. E., 2010, A$\&$A, 516, A106
\bibitem{latexcompanion}
Nespoli E. $\&$ Reig P., 2011, A$\&$A, 526, 7
\bibitem{latexcompanion}
Orlandini M., Dal Fiume D., Frontera F. $\&$ et al. 1999, A$\&$A 349, L9
\bibitem{latexcompanion}
Orlandini M., Frontera F., Masetti N. $\&$ et al. 2012, AJ, 748, 86
\bibitem{latexcompanion}
Press W. H., Teukolsky S. A., Vetterling W. T. $\&$ et al. 2007,
\bibitem{latexcompanion}
Numerical recipes. The art of scientific computing (Cambridge University Press)
\bibitem{latexcompanion}
Piraino S., Santangelo A., Segreto S. et al. 2000, A$\&$A, 357, 501
\bibitem{latexcompanion}
Pottschmidt K., Suchy S., Rivers E. et al. 2011, AIP Conf. Ser. 1427, 60
\bibitem{latexcompanion}
Pottschmidt K., Kreykenbohm I., Wilms J. $\&$ et al. 2005, ApJ, 634, 97
\bibitem{latexcompanion}
Reig P., Negueruela I., Buckley D. A. $\&$ et al. 2001, A$\&$A, 367, 266
\bibitem{latexcompanion}
Rivers E., Markowitz A., Pottschmidt K. et al. 2010, ApJ, 709, 179
\bibitem{latexcompanion}
Robba N. R., Burderi L., Di Salvo T. $\&$ et al. 2001, ApJ, 562, 950
\bibitem{latexcompanion}
Rodes-Roca J., Torrejon J., Kreykenbohm I. et al. 2009, A$\&$A, 508, 395
\bibitem{latexcompanion}
\bibitem{latexcompanion}
Shakura N., Postnov K., Kochetkova A., $\&$ Hjalmarsdotter L., 2012, MNRAS 420, 216
\bibitem{latexcompanion}
Santangelo A., del Sordo S., Segreto A. $\&$ et al. 1998, ApJ, 340, 55
\bibitem{latexcompanion}
Sakamoto T., Barthelmy S. D., Baumgartner W. et al. 2010, ATel, 2815
\bibitem{latexcompanion}
Schanne S., Gotz D., Grard L. et al. 2007, Proc. 6th Integral Workshop, 622, 479
\bibitem{latexcompanion}
Sguera V. Ducci L., Sidoliet L. $\&$ et al. 2010, MNRAS, 402, L49
\bibitem{latexcompanion}
Sguera V. Bazzano A., Ubertini P. $\&$ et al. 2014, ATel, 6664, 1S
\bibitem{latexcompanion}
Taani A., Zhang C.M., Al-Wardat M. $\&$ et al. 2012, Astronomische Nachrichten, 333, 53
\bibitem{latexcompanion}
Taani A., Al-Wardat M., $\&$ Zhao Y. H., 2013, International Journal of Modern Physics Conference Series, 23, 284
\bibitem{latexcompanion}
Taani A., 2016, Research in Astronomy and Astrophysics, 16, 101
\bibitem{latexcompanion}
Taani A. $\&$ Khasawneh A., 2017, Journal of Physics: Conference Series, 869, 012090
\bibitem{latexcompanion}
Taani A., Karino S., Song L. $\&$ et al. 2019a, Research in Astronomy and Astrophysics, 19, 12
\bibitem{latexcompanion}
Taani A., Karino S., Song L. $\&$ et al. 2019b, Journal of Physics: Conference Series, 1258, 012029
\bibitem{latexcompanion}
Terada Y., Mihara T., Nakajima M. $\&$ et al. 2006, ApJ, 648, 139
\bibitem{latexcompanion}
Tendulkar S. P., Furst F., Pottschmidt K. et al., 2014, ApJ, 795, 154
\bibitem{latexcompanion}
Truemper J., Pietsch W., Reppin C. et al. 1978, ApJ, 219, 105
\bibitem{latexcompanion}
Tsygankov S., Krivonos R. $\&$ Lutovinov A., 2012, MNRAS, 421, 2407T
\bibitem{latexcompanion}
Tsygankov S., Lutovinov A. $\&$ Krivonos R. $\&$ et al. 2015, MNRAS, 457, 258T
\bibitem{latexcompanion}
Tuerler M., Chenevez J., Bozzo E. et al., 2012, ATel, 3947
\bibitem{latexcompanion}
van den Heuvel E. P. J. 2004, Science, 303, 1143
\bibitem{latexcompanion}
Voges W., Pietsch W., Reppin C. et al. 1982, 263, 803
\bibitem{latexcompanion}
Wang W., 2009, MNRAS, 398, 1428
\bibitem{latexcompanion}
Wasserman I. $\&$ Shapiro S. L., 1983, Astrophy. J.265, 1036
\bibitem{latexcompanion}
Wei Y. C., Taani A., Pan Y. Y. et al., 2010, Chin. Phys. Lett., 27, 9801
\bibitem{latexcompanion}
Wheaton W., Doty J., Primini F. et al. 1979, Nature, 282, 240
\bibitem{latexcompanion}
Wilson C. A. $\&$ Finger M. H. 2005, ATel, 605
\bibitem{latexcompanion}
Wilson C., Finger M. $\&$ Camero-Arranz A., 2008 ApJ, 678, 1263
\bibitem{latexcompanion}
Wilms, J., 2006, AIPC, 840, 40
\bibitem{latexcompanion}
Yamamoto T., Sugizaki M., Mihara T. et al. 2011, Astron. Soc. Jpn. 63, 751
\end{theunbibliography}
\end{document}